\documentclass[aip,jcp,reprint]{revtex4-1}

\usepackage[T1]{fontenc}
\usepackage[utf8x]{inputenc}

\usepackage{amsmath}
\usepackage{amsfonts}
\usepackage{amssymb}

\DeclareMathOperator{\sgn}{sgn}
\usepackage[normalem]{ulem}

\usepackage{graphicx}

\usepackage[usenames]{color}
\usepackage{colortbl}
\usepackage{color}

\usepackage[verbose]{placeins}

\begin{document}


\title{Coarse-grained electrostatic interactions of coronene: Towards the crystalline phase}



\author{Thomas Heinemann}
\email[]{thomas.heinemann@tu-berlin.de}
\affiliation{Institut für Theoretische Physik, Technische Universität Berlin, Hardenbergstr. 36, 10623 Berlin, Germany}
\author{Karol Palczynski}
\email[]{karol.palczynski@helmholtz-berlin.de}
\affiliation{Institut für Physik, Humboldt Universität zu Berlin, Newtonstraße 15, 12489 Berlin, Germany}
\affiliation{Institut f{\"u}r Weiche Materie und funktionale Materialen, Helmholtz-Zentrum Berlin f{\"u}r Materialien und Energie, Hahn-Meitner Platz 1, 14109 Berlin, Germany}
\author{Joachim Dzubiella}
\email[]{joachim.dzubiella@helmholtz-berlin.de}
\affiliation{Institut für Physik, Humboldt Universität zu Berlin, Newtonstraße 15, 12489 Berlin, Germany}
\affiliation{Institut f{\"u}r Weiche Materie und funktionale Materialen, Helmholtz-Zentrum Berlin f{\"u}r Materialien und Energie, Hahn-Meitner Platz 1, 14109 Berlin, Germany}
\author{Sabine H. L. Klapp}
\email[]{klapp@physik.tu-berlin.de}
\affiliation{Institut für Theoretische Physik, Technische Universität Berlin, Hardenbergstr. 36, 10623 Berlin, Germany}




\begin{abstract}
In this article we present and compare two different, coarse-grained approaches to model electrostatic interactions of disc-shaped aromatic molecules, specifically coronene. Our study builds on previous work [J. Chem. Phys. \textbf{141}, 214110 (2014)] where we proposed, based on a systematic coarse-graining procedure starting from the atomistic level, an anisotropic effective (Gay-Berne-like) potential capable of describing van-der-Waals contributions to the interaction energy. To take into account electrostatics, we introduce, first, a linear quadrupole moment along the symmetry axis of the coronene disc. The second approach takes into account the fact that the partial charges within the molecules are distributed in a ring-like fashion. We then reparametrize the effective Gay-Berne-like potential such that it matches, at short distances, the ring-ring potential. To investigate the validity of these two approaches, we perform many-particle Molecular Dynamics (MD) simulations, focusing on the crystalline phase (karpatite) where electrostatic interaction effects are expected to be particularly relevant for the formation of tilted stacked columns. Specifically, we investigate various structural parameters as well as the melting transition. 
We find that the second approach yields consistent results with those from experiments despite the fact that the underlying potential decays with the wrong distance dependence at large molecule separations.
Our strategy can be transferred to a broader class of molecules, such as benzene or hexabenzocoronene.
\end{abstract}


\maketitle



\section{Introduction}

In the present article we analyze two methods to effectively incorporate electrostatics in symmetric disc-shaped molecules via an angle and temperature dependent coarse-grained potential.
The general purpose is to provide an approach that is appropriate to describe not only fluid phases, but also stable crystalline phases for these molecules.
Our work builds upon a previous investigation~\cite{firstpub2014} providing a general coarse-grained methodology for uniaxial discotics.
In the present work, we show that the relevant electrostatics can be treated separately from the remaining interactions via a direct sum.

In the literature there are many different approaches to create coarse-grained molecular models for pair interactions of discotics with electrostatics based on a Gay-Berne model~\cite{Golubkov2006,Bates1998,Neal1999,Neal1995,Neal2000} or different interaction potentials.~\cite{Miller1984,Babadi2006,  Schroer2001}
The simplest approach involves only one interaction site that inherits no orientational degree of freedom located at the molecules' center-of-mass.
For fluid phases at low densities, this kind of representation might indeed be sufficient, at least for molecules without complex internal structure.~\cite{Mognetti2008}
However, when considering systems at higher densities, the specific atomistic structure of the molecules yielding various degrees freedom becomes more and more important.
More specifically, at the other end of coarse-graining (before using pure quantum mechanics) are the atomistically-resolved models. Atomistic model studies for discotics have been proposed e.g. for coronene,~\cite{Rubio1996,Rapacioli2005,Fedorov2012} benzene~\cite{Levitt1988} or hexabenzororonene derivatives.~\cite{Andrienko2006}
In our investigation we make a compromise between both levels of detail.

A molecular orientation vector (uniaxial) or tensor (biaxial particle) is a frequently used example to account for orientational internal degrees of freedom.
Such an approach also allows to model anisotropic shape. Corresponding examples are the Gaussian-overlap~\cite{Berne1972} or the Gay-Berne potential~\cite{Gay1981} as well as their derivatives.~\cite{Kabadi1986,Everaers2003,firstpub2014}
Even more complexity can be reached by considering many (instead of one central) interaction sites.~\cite{Wilson1997,McBride1999}
Further, the treatment of electrostatics can be realized through an electric multipole attached to the interaction site in a centered~\cite{Miller1984,Golubkov2006,Schroer2001} or off-centered~\cite{Golubkov2008,Xu2013first} arrangement.
Moreover, by optimizing the arrangement of different electric multipoles per interaction site~\cite{Shen2014,Sokolovskii2009} one comes to more and more realistic models.~\cite{Gramada2011,Anandakrishnan2013}
A quite popular coarse-graining strategy is given by the Martini force-field~\cite{Marrink2007,Voth2008} which is not based on atoms but on chemical building blocks.

Here we rather aim at developing a model which has as few degrees of freedom as possible, while still conserving the uniaxiality and head-to-tail symmetry of typical disc-shaped molecules.

\begin{figure}[htb]
\begin{center}
   \includegraphics[height=0.15\textheight]{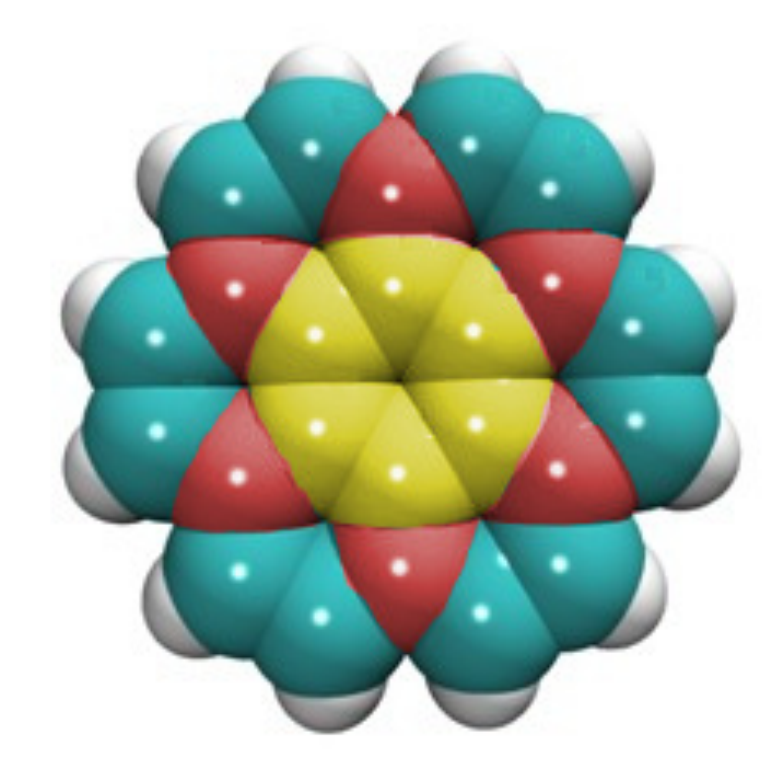}
   \end{center}
  \caption{\label{fig:COR}Sketch of a coronene molecule. The outer twelve small spheres represent hydrogen atoms while the remaining spheres represent carbon atoms. Equal color means equal distance from the center and equal charge contribution.
  }
\end{figure}

Throughout the entire analysis, we focus on coronene molecules (see Fig.~\ref{fig:COR}), for which we already developed an angle- and temperature dependent model for the van der Waals part with desired symmetries.~\cite{firstpub2014}
Various studies suggest a broad range of applications for coronene, which are outlined in the following.
It can be used as a building block for graphene nanoribbons,~\cite{Aguiar2014} as a candidate for active layer compounds in photovoltaic applications~\cite{Blumstengel2008} and its
derivatives~\cite{Choi2012,Rieger2008,Sanyal2013} can also be used in liquid crystal displays.~\cite{Choi2012}
Understanding coronene pair interactions gives further insight to graphene stacking~\cite{Ruuska2001,Meyer2003,Collignon2006} and growth of graphene.~\cite{Lloyd-Williams2012}
Even more, the crystal structure gives a relation to electronic properties, like band gaps.~\cite{Schatschneider2014}

We now turn to the modeling approach for coronene.
The angle- and temperature-dependent van der Waals model,~\cite{firstpub2014} on which our investigation is based on, stems from a coarse-graining procedure that uses atomistic trajectory data from two-molecule simulations (modeled with the generalized Amber force-field (GAFF)~\cite{Wang2004}).
This force-field was already successfully used in growth studies of another conjugated organic molecule, that is, para-sexyphenyl.~\cite{Palczynski2014}
It was shown that the force field yields the correct phase behavior (for a more detailed discussion of the validation of the Amber force-field for aromatics, see Olivier {et al.}~\cite{Olivier2014}).
Other coarse-grained potentials developed in the literature~\cite{Miller1984,Lilienfeld2006,Babadi2006,Obolensky2007} do not take into account the full angle- and temperature dependence.
In principle, the kind of coarse-graining method used in Ref.~\onlinecite{firstpub2014} could be applied to molecules with electrostatic contributions, when ab-initio simulations are used.
However, ab initio simulations are from a computational prospective very time consuming compared to atomistically-detailed simulations.
A common compromise consists in using static partial charges distributed among the atoms of the underlying microscopic model in the coarse-graining procedure.~\cite{Babadi2006}
By implication, as shown in previous studies involving quantum-chemical calculations,~\cite{Fedorov2012} it is generally not sufficient to use the static atomic partial charges characterizing an isolated dimer (except for selected
configurations such as a parallel displaced one~\cite{Zhao2008}).
However, for static partial charges an angle and temperature dependent coarse-grained pair potential for perylene, a flat but non-discotic molecule, was already developed by Babadi et al.~\cite{Babadi2006} using constrained steered dynamics for specific configurations desired
to fit an ellipsoidal soft-potential.~\cite{Everaers2003}
They also present a non-temperature dependent and biaxial model for coronene which implicitly involves static partial charges.

An elegant approach for the electrostatics in coronene is presented by Obolensky et al.~\cite{Obolensky2007}
They propose a uniaxial model for coronene consisting of concentric charged rings.
Unfortunately, this model is not temperature dependent, it is based on static partial charges and the evaluation of the interaction potential is quite involved due to numerical integrations for each pair interaction, inconvenient
for many-particle simulations.
Nonetheless, we also consider this electrostatic contribution, which already has the desired symmetries, namely uniaxiality and head-to-tail symmetry.

In particular, the raw model for our investigations is an additive combination of the temperature-dependent van der Waals potential from Ref.~\onlinecite{firstpub2014} and a coarse-grained electrostatic potential.
The entire temperature dependence stems from the van der Waals part alone.
The model thus implies that the impact of slight changes in the charge distribution on the interatomic forces is small against the van der Waals forces which dominate at short length scales.
If these temperature effects on the charge distribution are not important, we can focus on ground state charge distributions with the desired symmetry.

We introduce two different approaches to include the electrostatics in a simple model.
The first approach uses the van der Waals model and considers electrostatic interactions via a linear point quadrupole at the molecular center along the symmetry axis.
This kind of approach was already  applied for coronene~\cite{Miller1984} where an unusual strong repulsion for planar configurations was observed.
We address this issue in more detail later in our work.
Nonetheless, this approach was successfully applied to benzene molecules~\cite{Golubkov2006} in the liquid phase with an additional dampening field for closed distances.
Similar ideas are also used to model the interaction of clay particles,~\cite{Dijkstra1995,Trizac2002} which exhibit an electric double layer.

In the second approach, we include the pure ring electrostatics, suggested by Obolensky et al,~\cite{Obolensky2007} implicitly in our van der Waals model.

We deliberately do not use any kind of point charge electrostatic implementations in the models since point charge patterns overestimate the charge localization, are long-ranged and do not fulfill our symmetry requirements.

Both of the present approaches allow a better representation of the ${\pi-\pi}$ stacking~\cite{Hunter1990,Tsuzuki2005} which is also observed in similar aromatic molecules, e.g. benzene dimers~\cite{Tsuzuki2002,Tsuzuki2005} and hexabenzocoronene crystals.~\cite{Robertson1961}
\par
The remainder of this article is organized as follows. 
In section~\ref{sec:Models for pair interaction}, we introduce our models for further investigation.
Later on, the two model approaches are defined in Sec.~\ref{sec:Point quadrupole approach} and ~\ref{sec:Charged ring approach}.
These models are analyzed in Sec.~\ref{sec:Numerical analysis of the models} and used for many-particle simulations in Sec.~\ref{sec:Many-particle simulations} at room temperature (Sec.~\ref{sec:Bulk crystal at room temperature}) and beyond (Sec.~\ref{sec:Melting of bulk crystal}).
Finally, in Sec.~\ref{sec:Conclusion} we summarize our findings.

\section{\label{sec:Models for pair interaction}Models for pair interaction}
\subsection{General idea}

\begin{figure}[htb]
 \begin{center}
  \includegraphics[height=0.13\textheight]{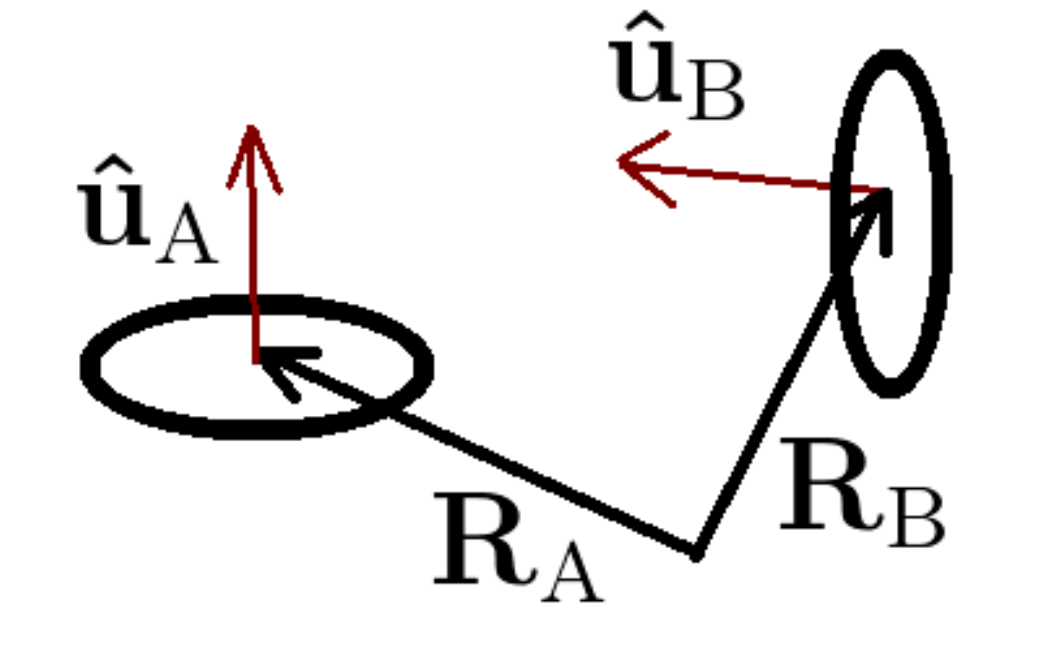}
 \end{center}
 \caption{\label{fig:gropairtiltednumbers}Sketch of a discotic dimer system with its corresponding reaction coordinates.}
\end{figure}

In the present work we assume additivity of the van der Waals and electrostatic contributions to the coronene-coronene interactions.
Thus, the full pair potential is given through
\begin{multline}
 U(\mathbf{R}_{\rm A},\mathbf{R}_{\rm B},\mathbf{\hat{u}}_{\rm A},\mathbf{\hat{u}}_{\rm B})=U_{\rm vdW}(\mathbf{R}_{\rm A},\mathbf{R}_{\rm B},\mathbf{\hat{u}}_{\rm A},\mathbf{\hat{u}}_{\rm B}) \\+U_{\rm elec.}(\mathbf{R}_{\rm A},\mathbf{R}_{\rm B},\mathbf{\hat{u}}_{\rm A},\mathbf{\hat{u}}_{\rm B})\text{,}
\end{multline}
where $U_{\rm vdW}$ and $U_{\rm elec.}$ are the van der Waals and electrostatic parts, respectively.
These pair potentials are described as functions of the center of mass positions $\mathbf{R}_{\rm A}$, $\mathbf{R}_{\rm B}$ and the axial
orientations $\mathbf{\hat{u}}_{\rm A}$ and $\mathbf{\hat{u}}_{\rm B}$.
A configuration showing two discotics with these vectorial reaction coordinates is presented in Fig.~\ref{fig:gropairtiltednumbers}. 
This set can be further reduced to four scalar reaction coordinates due to our symmetry requirements. These are, besides translational and rotational symmetry, uniaxial and head-to-tail symmetry of the molecules (see Appendix~\ref{sec:Dimer configurations}). The reduced set of reaction coordinates is important for calculating configuration dependent histograms (see Ref.~\onlinecite{firstpub2014}).
For the following investigations, it is more appropriate to use the vector-based reaction coordinates.

\subsection{\label{sec:Treatment of the van der Waals interaction}Treatment of the van der Waals interaction}
A coarse-grained model describing van der Waals interactions has been developed in Ref.~\onlinecite{firstpub2014}.
Starting from an atomistic molecular model, involving all non-electrostatic interactions, we have numerically calculated the potential of mean force for several
pair configurations and temperatures. Then, we have parametrized the resulting curves using a modified Gay-Berne potential ($U_{\rm mGB}$).
The corresponding parameters are given in Appendix ~\ref{sec:Parametrizations} (so-called ``vdW parametrization'').
To sum up, we treat the van der Waals interaction using the potential
\begin{align}
U_{\rm vdW}(\mathbf{R},\mathbf{\hat{u}}_{\rm A},\mathbf{\hat{u}}_{\rm B})&\approx U_{\rm mGB}(\mathbf{R},\mathbf{\hat{u}}_{\rm A},\mathbf{\hat{u}}_{\rm B}) \bigg| \begin{array}{l}{\textrm{\tiny vdW}}\\{\textrm{\tiny parametrization}}\end{array}\nonumber\\
&=U_{\rm mGB}^{\rm vdW}(\mathbf{R},\mathbf{\hat{u}}_{\rm A},\mathbf{\hat{u}}_{\rm B})\text{.}
\label{eqn:vdW potential}
\end{align}
We stress that this parametrization is indeed temperature-dependent due to the rigorous coarse-graining procedure described in Ref.~\onlinecite{firstpub2014}.
Hereby $\mathbf{R}=\mathbf{R}_{\rm B}-\mathbf{R}_{\rm A}$ represents the intermolecular connecting vector.
The mGB-potential is uniaxial in symmetry (suggested by the $D_{6h}$-symmetry of coronene) and reads
\begin{multline}
U_{\rm mGB}(\mathbf{R},\mathbf{\hat{u}}_{\rm A},\mathbf{\hat{u}}_{\rm B})=4 \, \epsilon(\mathbf{\hat{R}},\mathbf{\hat{u}}_{\rm A},\mathbf{\hat{u}}_{\rm B}) \times \\ \times \left[ \left(\frac{1}{R^*}\right)^{12}- \left(\frac{1}{R^*}\right)^{6} \right]\text{.}
\label{eqn:UmGB}
\end{multline}
Here, ${R^*=(R-\sigma(\mathbf{\hat{R}},\mathbf{\hat{u}}_{\rm A},\mathbf{\hat{u}}_{\rm B})+d_{\text{w}} \sigma_0)/(d_{\text{w}} \, \sigma_0})$ represents the reduced distance
with the contact function
\begin{multline}
 \sigma(\mathbf{\hat{R}},\mathbf{\hat{u}}_{\rm A},\mathbf{\hat{u}}_{\rm B})= \\ \sigma_0 \left[1- \frac{\chi}{2}  \left( A^+ + A^- \right) + (1-\chi)\,\chi_{\text{t}} \left( A^+ \,A^- \right)^{\gamma} \right]^{-\frac{1}{2}}\text{,}
 \label{eqn:contact distance}
\end{multline}
the well width $d_{\text{w}} \, \sigma_0$ and the well depth $\epsilon$. The latter is given by
\begin{multline}
 \epsilon(\mathbf{R},\mathbf{\hat{u}}_{\rm A},\mathbf{\hat{u}}_{\rm B})=\epsilon_0 \, \left[1-\chi^2 (\mathbf{\hat{u}}_{\rm A}\cdot\mathbf{\hat{u}}_{\rm B})^2    \right]^{-\frac{1}{2}\,\nu} \times \\ \times
 \left[\epsilon_{\text{M}}(\mathbf{R},\mathbf{\hat{u}}_{\rm A},\mathbf{\hat{u}}_{\rm B})\right]^{\mu}\text{.}
 \label{eqn:epsilon}
\end{multline}
For the coefficients in Eq.~\eqref{eqn:contact distance}, we set $A{^{\pm}=((\mathbf{\hat{u}}_{\rm A}\cdot\mathbf{\hat{R}})\pm (\mathbf{\hat{u}}_{\rm B}\cdot\mathbf{\hat{R}}))^2/(1 \pm \chi (\mathbf{\hat{u}}_{\rm A}\cdot\mathbf{\hat{u}}_{\rm B}))}$ with the anisotropy parameters
${\chi=(\kappa^2-1)/(\kappa^2+1)}$ and ${\chi_{\text{t}}=\left[(\kappa-1)/(\kappa+1)\right]^2}$, where  ${\kappa=\sigma_{\text{FF}}/\sigma_0}$ is the quotient of the face-face and edge-edge contact distance.
Regarding the well depth $\epsilon$ [see Eq.~\eqref{eqn:epsilon}], we use the well-known GB formula,~\cite{Gay1981}
where the overlap factor $\epsilon_{\text{M}}$ is modified (as compared to the original definition~\cite{Gay1981}) according to
\begin{multline}
\label{eqn:epsilonM}
\epsilon_{\text{M}}(\mathbf{\hat{R}},\mathbf{\hat{u}}_{\rm A},\mathbf{\hat{u}}_{\rm B})=1- \frac{\chi'}{2}  \left( A'^+ + A'^- \right) + \theta \left( A'^+ \,A'^- \right)^{\gamma'} 
\\+ \xi\, K(\mathbf{\hat{R}},\mathbf{\hat{u}}_{\rm A},\mathbf{\hat{u}}_{\rm B})\text{.}
\end{multline}
The coefficients $A'^{\pm}$ in Eq.~\eqref{eqn:epsilonM} resemble the quantities $A^{\pm}$, but incorporate the anisotropy parameter $\chi'$. 
The function
\begin{multline}
\label{eqn:K}
K(\mathbf{\hat{R}},\mathbf{\hat{u}}_{\rm A},\mathbf{\hat{u}}_{\rm B})= 1-5(\mathbf{\hat{u}}_{\rm A}\cdot\mathbf{\hat{R}})-5(\mathbf{\hat{u}}_{\rm B}\cdot\mathbf{\hat{R}})\\-15 (\mathbf{\hat{u}}_{\rm A}\cdot\mathbf{\hat{R}})^2 (\mathbf{\hat{u}}_{\rm B}\cdot\mathbf{\hat{R}})^2 \\
+2\left((\mathbf{\hat{u}}_{\rm A}\cdot\mathbf{\hat{u}}_{\rm B})-5(\mathbf{\hat{u}}_{\rm A}\cdot\mathbf{\hat{R}})\, (\mathbf{\hat{u}}_{\rm B}\cdot\mathbf{\hat{R}})\right)^2 
\end{multline}
has the symmetry of a  linear quadrupole-quadrupole interaction.~\cite{Stone1978,Gray1984,Bates1998,Boublik1990}

\subsection{\label{sec:Electrostatic interaction}Electrostatic interaction}

To get a picture of the electrostatic properties characterizing coronene, we start from quantum-chemical results for the atomistic charge distribution given by Obolensky et al. (Ref.~\onlinecite{Obolensky2007}).
As seen from Fig.~\ref{fig:COR}, there are four different distances from the molecular center where the atoms are placed in a sixfold symmetry.
All atoms with the same distance are considered to carry the same partial charge.
We next focus on the spherical multipole decomposition of the electrostatic potential, that is~\cite{Gray1984}
\begin{align}
\Phi(r,\phi,\theta)= \frac{1}{4\pi\varepsilon_0} \sum_{\ell=0}^\infty \sum_{m=-\ell}^{\ell}\; \frac{\left(\frac{4\pi}{2 \ell+1}\right) Q_{\ell,m}}{r^{\ell+1}}Y_{\ell m}^*(\theta,\phi)\text{,}
\label{eqn:Phi}
\end{align}
where $Y_{\ell m}$ are the spherical harmonics depending on the polar angle $\theta$ and the azimuthal angle $\phi$.
The dielectric constant is denoted by $\varepsilon_0$, and the multipole moments are defined through
\begin{align}
 Q_{\ell,m}&=\sum_i q_i \left\|\mathbf{r}_i\right\|^\ell \, Y_{\ell m}(\theta_i,\phi_i)
\text{.}
\end{align}
Several symmetries in the charge distribution effectively lead to a reduction of the number of $Q_{\ell,m}$.
In particular, the net charge and the dipole moment of the proposed charge distribution are zero, i.e. $Q_{0,0}=0$ and $Q_{1,m}=0$.
The head-tail-symmetry ($\theta_i=\pi/2\rightarrow \pi-\theta_i$) reduces $l$-values to even numbers.
Summing up, our multipole decomposition of the electrostatic potential consists of a quadrupole moment ($l=2$) and higher multipoles with $l \in (4,6,8, \dots)$.

For molecules with continuous axial symmetry, we would only have to consider multipole moments with $m=0$.
However, the underlying atomistic model of coronene (see Fig.~\ref{fig:COR}) does not have full (continuous) axial symmetry.
Nonetheless, the sixfold symmetry suggests that the linear quadrupole moment ($Q_{2,0}$) is the only non-zero quadrupole moment.
That means the next non-vanishing multipole contribution after the linear quadrupole is the linear hexadecapole ($Q_{4,0}$).
We next consider the quotient of the coefficients in the multipole expansion corresponding to $Q_{2,0}$ and $Q_{4,0}$ [see Eq.~\eqref{eqn:Phi}] to evaluate the possibility of neglecting higher multipole terms ($\ell=4$ and beyond). Specifically, we consider
\begin{align}
 {\mathcal Q}=\frac{Q_{\ell,m}/(2\ell+1)/r^{\ell+1}}{Q_{\bar{\ell},\bar{m}}/(2\bar{\ell}+1)/r^{\bar{\ell}+1}}\bigg|_{\ell=2,m=0,\bar{\ell}=4,\bar{m}=0}\text{.}
\end{align}
At large distances, such as $r\ge 2.4\,{\rm nm}$, the absolute value of ${\mathcal Q}$  is beyond 30, i.e. the linear quadrupole
approach is reasonable.
However, for distances smaller than $r=0.44\,{\rm nm}$ (a typical distance in the crystalline phase) the quotient ${\mathcal Q}$ produces values below unity.
Thus, the quadrupole approximation is no more eligible.
To somewhat compensate this effect we here consider also a small quadrupole interaction strength to avoid an overestimation of the bonding.

\subsubsection{\label{sec:Point quadrupole approach}Point quadrupole approach}

In this approach the entire electrostatics is represented by a point quadrupole tensor.
This approach is aimed at maintaining the electrostatics in the far field.
The quadrupole tensor is given by~\cite{Gray1984}
\begin{align}
 \mathbf{Q}=\frac{1}{2}\sum_{\rm i} q_i \, \left[ 3\,\mathbf{r}_i\otimes\mathbf{r}_i -(\mathbf{r}_i)^2\mathbf{1}  \right]\text{,}
\end{align}
where $q_i$ is the charge of atom $i$ placed at $\mathbf{r}_i$ from the geometric center.
In the eigenbasis $\mathbf{P}$, the quadrupole tensor of a uniaxial charge distribution has the following symmetry
\begin{align}
\label{eqn:diagonal P tensor}
 \mathbf{P}\mathbf{Q}\mathbf{P}^{-1}=Q\left[ \begin{array}{ccc}
                -1/2&0&0\\0&-1/2&0\\0&0&1\\
               \end{array}
\right]\text{.}
\end{align}
Thus, the quadrupole tensor can be written in terms of a linear quadrupole pointing along the molecular symmetry axis.
The prefactor in Eq.~\eqref{eqn:diagonal P tensor}, $Q=\sqrt{4\pi/5}\,Q_{2,0}$, marks the quadrupole strength of the linear quadrupole.~\cite{Gray1984}

At this point, we can define the ``point quadrupole model'' (pq model) through the following additive formula
\begin{multline}
  U_{\text{pq}}(\mathbf{R},\mathbf{\hat{u}}_{\rm A},\mathbf{\hat{u}}_{\rm B})=U_{\rm mGB}^{\rm vdW}(\mathbf{R},\mathbf{\hat{u}}_{\rm A},\mathbf{\hat{u}}_{\rm B})\\+\frac{1}{4\pi\varepsilon_0}\frac{3}{4} \frac{Q^2}{R^5}\cdot  K(\mathbf{\hat{R}},\mathbf{\hat{u}}_{\rm A},\mathbf{\hat{u}}_{\rm B})\text{,}
  \label{eqn:Upointquad}
\end{multline}
where the second term on the right side describes the energy of one linear quadrupole in the field of another linear quadrupole of equal strength $Q$, and the function $K$ is defined in Eq.~\eqref{eqn:K}.

For simplicity, we henceforth use the dimensionless quadrupole moment
\begin{align}
 Q^*=\left\|\frac{Q}{\sqrt{4\pi\varepsilon_0 \, \mathrm{kJ}/\mathrm{mol}\,\mathrm{nm}^5}}\right\|\text{.}
\end{align}
The absolute is used because the sign does not influence the pair potential defined in Eq.~\eqref{eqn:Upointquad}.
The values of $Q^*$ resulting from density functional calculations performed in Ref.~\onlinecite{Obolensky2007} range from $0.617$ to $0.83$.
A well-known problem with a point multipole representation~\cite{Thole1981, Taylor2013, Golubkov2006} of the electrostatic interaction of extended molecules is its failure for closed distances, due to a spatially extended charge distribution and induced dipoles.
In particular, attractive configurations become infinitely strong for vanishing quadrupole-quadrupole distances.
Indeed, we will later see in Sec.~\ref{sec:Potential curves} that the latter effect leads to an overestimation of the binding energy for parallel displaced configurations.

We here consider quadrupole strengths that are in the range of the previously termed reference values.
Specifically, we use: ${Q^*=0, 0.5, 1, 1.5}$.
It is thus possible to gradually study the influence of the point quadrupole on the cohesion energy.

\subsubsection{\label{sec:Charged ring approach}Charged ring approach}
\begin{widetext}
The idea of this approach is to better take into account interactions for small distances.
The atomic partial charges in the atomistic model of Obolensky et al,~\cite{Obolensky2007} have the same distance to the geometrical center (see Fig.~\ref{fig:COR}) and are equally distributed along a ring.
This makes it plausible to describe each ring as uniformly charged, thus fulfilling the symmetry requirements of our coarse-grained potential (uniaxiality and head-to-tail symmetry).
We also note that considering a smeared charge distribution due to the rings is not necessarily a less accurate choice compared to an atomistic point charge distribution. Indeed, it seems plausible that mapping the quantum mechanical charge density onto atomic point charges can lead to a too strong localization.

For simplicity, we here consider (as proposed in Ref.~\onlinecite{Obolensky2007}) only the two outer rings for the electrostatics.
Hereby the hydrogen ring is charged with $Q_1=12\cdot 0.15 e$ and the next inner ring is oppositely charged, i.e.  $Q_2=-Q_1$.
The sum of all ring-ring interactions is thus given by
\begin{align}
 U_{\rm ring-ring}(\mathbf{R},\mathbf{\hat{u}}_{\rm A},\mathbf{\hat{u}}_{\rm B})&=\sum_{i}^{\rm rings\, of\, A}\,\sum_{j}^{\rm rings\, of\, B}U_{\textrm{ring $i$-ring $j$}}(\mathbf{R},\mathbf{\hat{u}}_{\rm A},\mathbf{\hat{u}}_{\rm B})\text{.}
 \label{eqn:ring-ring}
\end{align}
Hereby a single ring-ring interaction is given by a line integral,~\cite{Obolensky2007}
\begin{align}
U_{\textrm{ring $i$-ring $j$}}(\mathbf{R},\mathbf{\hat{u}}_{\rm A},\mathbf{\hat{u}}_{\rm B}) &=\frac{Q_j}{2\pi R_j}\oint_{\textrm{ring$j$}}\mathrm{d}\mathbf{r} \,\Phi(R_i,\sin(\theta),r)\text{,}
\end{align}
where the vector $\mathbf{r}=R \,\mathbf{\hat{R}}+ \,\uline{\mathcal{R}}(\mathbf{\hat{u}}_{\rm B})\left[\begin{array}{c} R_j \cos\phi \\ R_j \sin\phi \\0 \end{array}\right]$
points from the center of ring $i$ to a point on ring $j$.
The symbol $\uline{\mathcal{R}}$ marks a rotation matrix that rotates the x-y-plane orthogonal to $\mathbf{\hat{u}}_{\rm B}$, while $R_j$ and $\phi$ denote the polar coordinates of ring $j$.
The electrostatic potential exerted from ring $i$ with radius $R_i$ on a position, described by the relative spherical coordinates $\theta$ and $r$, is defined through
\begin{align}
\Phi(R_i,\sin(\theta),r)&=\frac{1}{4\pi\epsilon_{\rm pm}}\frac{Q_i}{2\pi} \cdot \left(\frac{2 \mathcal{K}\left[\frac{4 r R_i \sin(\theta)}{r^2+2 R_i \sin(\theta) r+R_i^2}\right]}{\sqrt{r^2+2 r R_i \sin(\theta)+R_i^2}}+\frac{2 \mathcal{K}\left[-\frac{4 r R_i \sin(\theta)}{r^2-2 R_i \sin(\theta) r+R_i^2}\right]}{\sqrt{r^2-2 r R_i \sin(\theta)+R_i^2}}\right)\text{,}
\end{align}
where  $\sin(\theta)=\left\|\mathbf{r}/r\times \mathbf{\hat{u}}_{\rm A}\right\|$ and $\mathcal{K}$ stands for the elliptic integral K.

Direct combination of the van der Waals potential Eq.~\eqref{eqn:vdW potential}] and the ring-ring potential [Eq.~\eqref{eqn:ring-ring}] yields
\begin{align}
  U_{\text{direct}}(\mathbf{R},\mathbf{\hat{u}}_{\rm A},\mathbf{\hat{u}}_{\rm B})=U_{\rm mGB}^{\rm vdW}(\mathbf{R},\mathbf{\hat{u}}_{\rm A},\mathbf{\hat{u}}_{\rm B})+U_{\textrm{ring-ring}}(\mathbf{R},\mathbf{\hat{u}}_{\rm A},\mathbf{\hat{u}}_{\rm B})\text{.}
  \label{eqn:U direct}
\end{align}
\end{widetext}
Henceforth, we term the latter potential as  ``direct'' potential.
We note that $U_{\text{direct}}$ is temperature-dependent as a consequence of the temperature-dependent parametrizations used for the van der Waals model $U_{\rm mGB}^{\rm vdW}$.
From a computational perspective, however, this kind of potential is not very suitable, at least not for many-particle simulations. This is,
one has to calculate ring integrals numerically for each pair of particles.
Therefore, we here propose a way to effectively include the ring-ring interactions from above by simply reparametrizing the modified Gay-Berne model used for the van der Waals interaction [see Eq.~\eqref{eqn:UmGB}] with new parameters chosen to match the ``direct'' potential [Eq.~\eqref{eqn:U direct}] for different angular arrangements.
This computationally more simple approach is aimed at correctly describing the direct potential in the near field.
We name it ``implicit electrostatics model''. Specifically,
\begin{align}
 U_{\rm implic.}(\mathbf{R},\mathbf{\hat{u}}_{\rm A},\mathbf{\hat{u}}_{\rm B})=U_{\rm mGB}(\mathbf{R},\mathbf{\hat{u}}_{\rm A},\mathbf{\hat{u}}_{\rm B})\bigg| \begin{array}{l}{\textrm{\tiny implic.\, electr.}}\\{\textrm{\tiny parametrization}}\end{array}\text{.}
 \label{eqn:U implic}
\end{align}
The consequences for the long-range behavior are discussed in Sec.~\ref{sec:Electrostatic long range behavior}.
In our implicit electrostatics parametrization, we fix the following parameters to the corresponding van der Waals values for all temperatures: ${\mu=1}$, ${\nu=1}$, ${\gamma=4}$ and ${\gamma'=4}$.
The remaining parameters are extracted from the direct potential [Eq.~\eqref{eqn:U direct}] using the same angular configurations as those used in Ref.~\onlinecite{firstpub2014} for the van der Waals interaction.
An overview of different pair configurations, including those needed to fit the implicit electrostatics potential [Eq.~\eqref{eqn:U implic}] with the direct potential [Eq.~\eqref{eqn:U direct}], is presented in Appendix~\ref{sec:Dimer configurations}.
The anisotropy parameter $\chi$ is calculated by measuring the face-face contact distance $\sigma_{\text{FF}}$ and the edge-edge contact distance $\sigma_0$.
The well width is calculated using $\sigma_{\text{FF}}$ and the distance corresponding to the minimum of the face-face potential, $R_{\text{FF}}^{\text{min}}$, yielding
\begin{align}
 d_{\text{w}}=\frac{R_{\text{FF}}^{\text{min}}-\sigma_{\text{FF}} }{\sigma_0 \, \left( 2^{1/6} -1 \right)}\text{.}
\end{align}
To summarize, the two parameters which determine the shape, are extracted from the face-face and edge-edge configuration.
Further configurations come into play when we determine the remaining parameters $\epsilon_0$, $\chi'$, $\theta$ and $\xi$ 
by fitting the results for $U_{\text{direct}}(\mathbf{R},\mathbf{\hat{u}}_{\rm A},\mathbf{\hat{u}}_{\rm B})$ according to Eq.~\eqref{eqn:epsilon}. Specifically,
our parameter fit builds on the four attractive wells stemming from the
parallel weakly displaced, parallel displaced, T and edge-edge configuration.
Finally, the specific parameters for all considered temperatures are given in Appendix ~\ref{sec:Parametrizations} (``implicit electrostatics parametrization'').

\section{\label{sec:Numerical analysis of the models}Numerical analysis of the models}
In this section we first analyze the potential curves from the different electrostatic approaches described in Sec.~\ref{sec:Electrostatic interaction}.
Second, we focus on the resulting long-range behavior.

\subsection{\label{sec:Potential curves}Potential curves}

In the following, we investigate the  potential curves resulting from Eq.~\eqref{eqn:Upointquad} (point quadrupole model), Eq.~\eqref{eqn:U direct} (direct model) and Eq.~\eqref{eqn:U implic} (implicit electrostatics model)
for different molecular configurations at a temperature of $T=300$ \,K.

\begin{figure}[htb]
   \includegraphics[width=8.5cm]{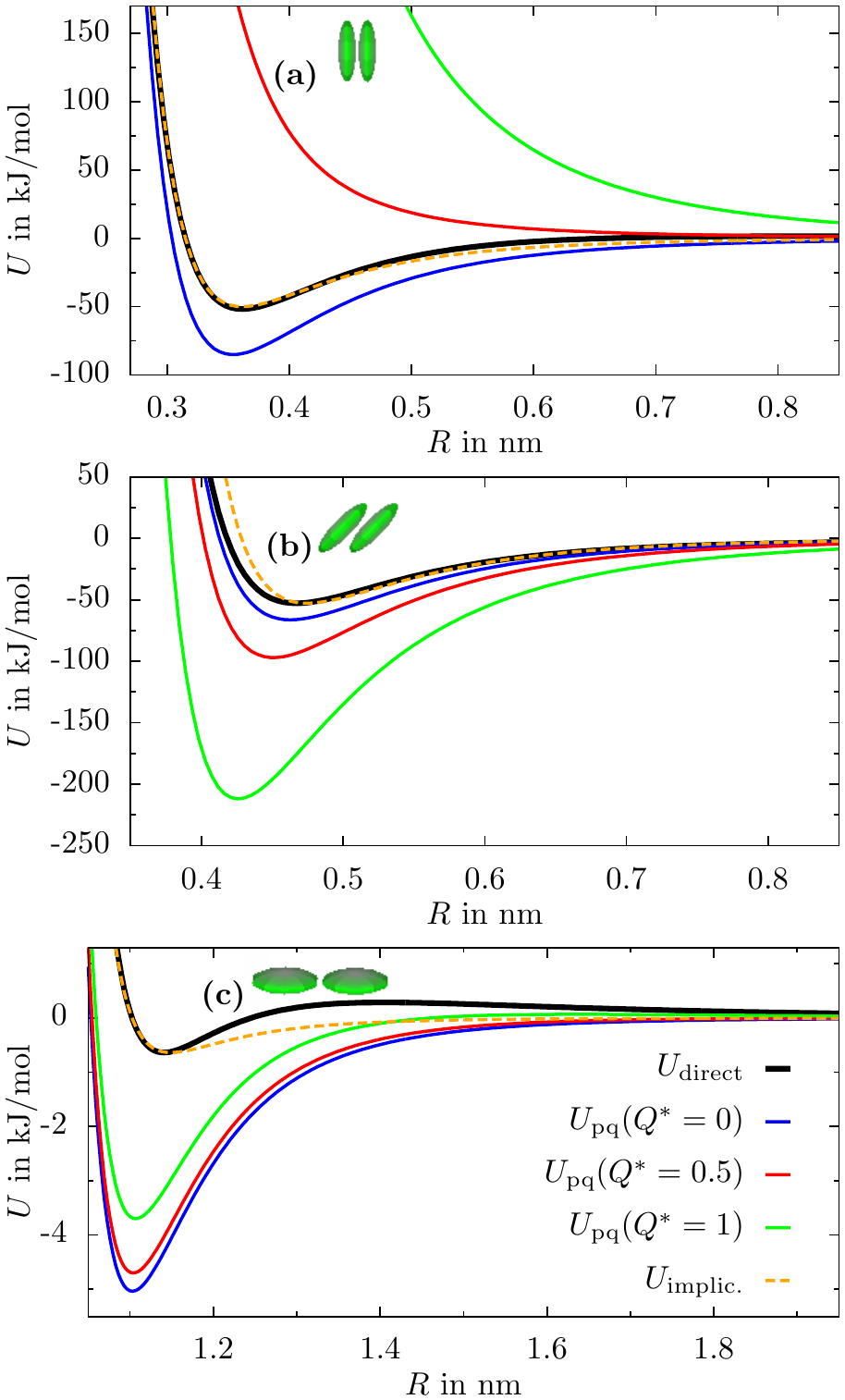}
    \caption{\label{fig:differentq} 
    Potential curves along the intermolecular distance of the direct model, point quadrupole model (with reduced strengths: $0$, $0.5$, $1$) and the implicit electrostatics model at $300$\,K for fixed configurations. The point quadrupole model with ${Q^*\!=\!0}$ coincides per definition with the van der Waals model. 
    }
\end{figure}

In Fig.~\ref{fig:differentq}, these potential curves are presented for three molecular pair configurations.

Of special interest are the corresponding energy minima.
For background information, Rapacioli et al.~\cite{Rapacioli2005}, who performed ground state calculations of coronene, found similar values for the binding energy in the face-face (ff)
configuration ($-94.9\,{\rm kJ}/{\rm mol}$ for tooth-to-tooth alignment and $-92.35\,{\rm kJ}/{\rm mol}$ for a $30^{\circ}$\, twisted setup) and the parallel displaced (pd) configuration ($-93.66\,{\rm kJ}/{\rm mol}$).

Similar numerical values are also found by Zhao et al.~\cite{Zhao2008} except for the perfectly stacked face-face configuration (``tooth-to-tooth'' setup), which is only half of the magnitude than for twisted stacking in their calculations.
Our present analysis shows that the direct model is characterized by a similar binding energy for the ff-configuration ($-50.02\,{\rm kJ}/{\rm mol}$) in comparison with the pd-configuration ($-52.778\,{\rm kJ}/{\rm mol}$) as observed in the work of
Rapacioli et al.~\cite{Rapacioli2005}
However, the magnitude is only about half of that obtained in Ref.~\onlinecite{Rapacioli2005}.
Besides the fact we consider a finite temperature of $300$\,K and use different electrostatics implementations, differences also arise from the fact that our underlying molecular mechanics parameters are taken from the generalized AMBER force field~\cite{Wang2004} while in Ref.~\onlinecite{Rapacioli2005} parameters from van de Waal were used.
A figure containing also ground state investigations is presented in Appendix~\ref{sec:Ground state curves}.
Moreover a short analysis of the temperature influence on the direct model in the face-face configuration can be found in Appendix~\ref{sec:Temperature dependence}.

Anyhow, when comparing potential curves for the implicit electrostatics model and the direct model at $300$\,K, we find a good agreement.
This is because the implicit electrostatics model is designed to fit the direct model in closed contact configurations.

Concerning the point quadrupole model, we notice a strong repulsion for the ff-con\-figuration as depicted in Fig.~\ref{fig:differentq}(a), even at the smallest considered non-vanishing quadrupole strength of $Q^*=0.5$. 
This behavior was already observed in Ref.~\onlinecite{Miller1984}.
In order to reach the potential depth of the direct model in this configuration, the quadrupole strength should be very close to zero, which implies a weak far-field behavior.
Moreover, even the smallest quadrupole strength further increases the potential depth of the pd-configuration [see Fig.~\ref{fig:differentq}(b)] away from its reference value.
Considering with $Q^*=0.5$ only a fraction of the quadrupole strength (reference values~\cite{Obolensky2007} between $Q^*=0.617$\, to\, $0.83$ for various charge distributions; $Q^*=0.963$ for the proposed two-ring model), leads to a two times stronger potential depth for the pd-configuration compared to the ff-configuration of the direct model.
This is not consistent with ground state results.~\cite{Rapacioli2005,Zhao2008} 
Nonetheless, we want to investigate in Sec.~\ref{sec:Many-particle simulations}, whether a point quadrupole model is still able to predict the correct melting behavior of the coronene crystal.

In Fig.~\ref{fig:differentq}(c), the potential curves of the edge-edge configuration reveal a strong influence of the electrostatics at all distances. This becomes clear when comparing $U_{\rm direct}$ with the pair potential of the van der Waals model
$U^{\rm vdW}_{\rm mGB}=U_{\rm pq}(Q^*=0)$.
The edge-edge configuration is in part responsible for the binding between columnar arrangements in a crystal.

\subsection{\label{sec:Electrostatic long range behavior}Electrostatic long-range behavior}

The implicit electrostatics model [see Eq.~\eqref{eqn:U implic}], which seems most promising so far, decays with the center-of-mass distance $R$ as $R^{-6}$.
This means that (contrary to the point quadrupole model) the implicit electrostatics model cannot reproduce the long-range behavior expected by the electrostatic interactions, which is $R^{-5}$ (because the quadrupole is the smallest non-vanishing multipole; see Sec. ~\ref{sec:Electrostatic interaction}).
We note that, the ring-ring potential [see Eq.~\eqref{eqn:ring-ring}], which is the underlying electrostatics for the implicit electrostatics model, reveals a correct $R^{-5}$ decay.
How does this ring-ring potential decay at short and intermediate distances, which are particularly relevant in the present context?
To this end we consider the function
\begin{align}
 S(R)=\frac{\partial \log(U_{\rm ring-ring}(\mathbf{R},\mathbf{\hat{u}}_{\text{A}},\mathbf{\hat{u}}_{\text{B}}))}{\partial \log(R)}\text{.}
\end{align}
This function is depicted in Fig.~\ref{fig:exponentialslope} for various configurations (see Appendix~\ref{sec:Dimer configurations}).
\begin{figure}[htb] 
 \begin{center}
  \includegraphics[width=8.5cm]{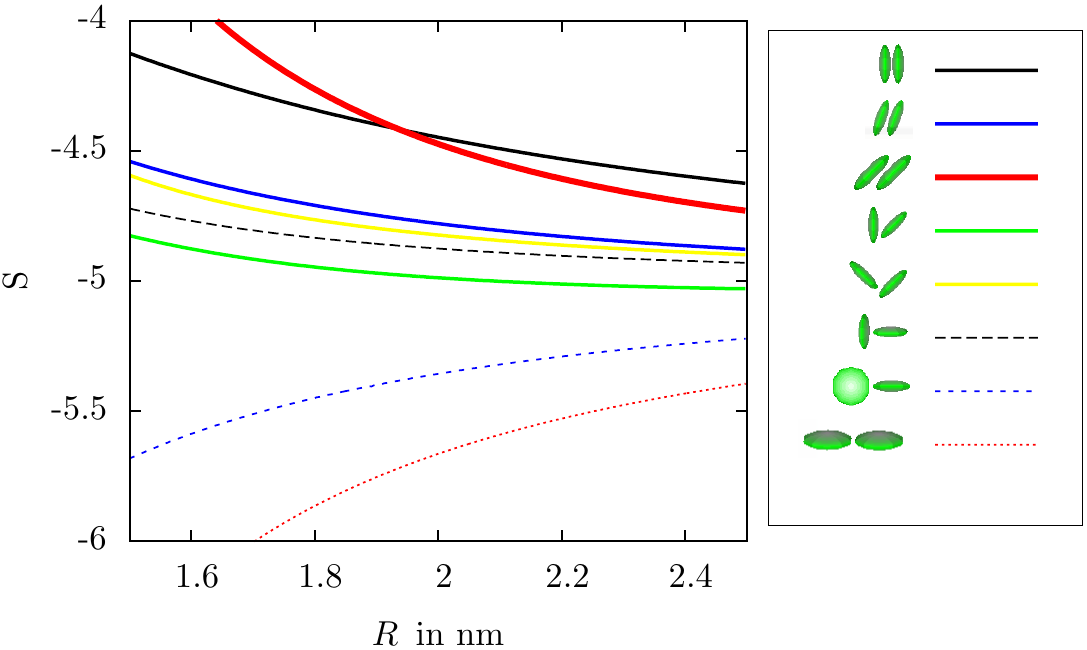}
 \end{center}
 \caption{\label{fig:exponentialslope}Power-law exponent of the ring-ring potential for several molecular configurations as function of $R$.
 }
\end{figure}
It turns out that $S(R)$ is not approaching the far-field limit of $-5$ in the range of interest, that is, $R<2.4$\,nm. Therefore it is sufficient 
to model the electrostatics in our range of interest implicitly with a van der Waals model revealing a slightly different long-range behavior.
A similar strategy was used to treat the electrostatics and the van der Waals interactions in DNA.~\cite{MorrisAndrews2010}

\section{\label{sec:Many-particle simulations}Many-particle simulations}

In the current section we investigate a many-particle system of coronene molecules using molecular dynamics at constant temperature (T) and pressure (P).
First, we investigate the equilibrium structural properties of the unit cell at room temperature.
Second, we analyze how certain structural parameters behave upon heating up the crystal until it melts.
Each simulation run starts from lattice-like initial conditions (described below) and leads to a relaxation of particle positions and box-vectors.
To affect box lengths as well as box-angles we use anisotropic pressure coupling.
In particular, the temperature and pressure coupling was realized with the Berendsen weak coupling scheme,~\cite{Berendsen1984} also used in our previous work.~\cite{firstpub2014}
Rotational dynamics was solved in a similar fashion, as proposed by Fincham et al.~\cite{Fincham1993} with separate temperature control.

\begin{figure}[htb]
 \begin{center}
  \includegraphics[width=5.5cm]{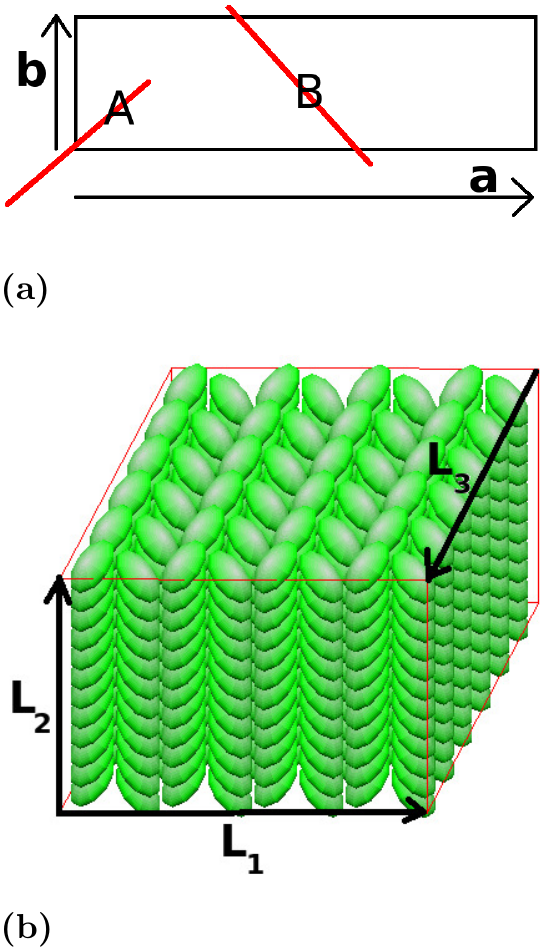}
 \end{center}
 \caption{\label{fig:unitcell}(a) Unit cell with molecules A and B (the red lines represent the side view of the discotic molecules). (b) Snapshot of a real crystalline configuration.
  }
\end{figure}

The coronene molecules are modeled with either the point quadrupole model (see Sec.~\ref{sec:Point quadrupole approach}) or the implicit electrostatics model (see Sec.~\ref{sec:Charged ring approach}).
Our many-particle system for all simulations consists of 576 molecules and is initialized with the following arrangement of unit cells in terms of box-vectors: $\mathbf{L}_1=4\, \mathbf{a}$, $\mathbf{L}_2=12\,\mathbf{b}$ and $\mathbf{L}_3=6\,\mathbf{c}$.
Hereby $\mathbf{a}$, $\mathbf{b}$ and $\mathbf{c}$ denote the right-handed unit cell vectors, which together with the molecular arrangement and orientation in the unit cell are taken from Ref.~\onlinecite{Robertson1945}.
The molecules are aligned in a so-called herringbone pattern, which resembles tractor traces when seen from the side (see Fig.~\ref{fig:unitcell}).

The pressure is set to $1$ bar and the compressibility to ${2.25\cdot 10^{-4} \,\mathrm{bar}^{-1}}$. Relaxation time constants involved in T- and P-control are set to ${2\,\mathrm{ps}}$.
The equilibration times range between $3.5$ and $100$ ns at a time step of $10$ fs. 
A force-shifted cutoff was used for the pair forces and set to $2.4$ \,nm.
To test this approach, we calculated the total electrostatic energy both, with the cutoff and by using the full Ewald sum for quadrupoles
(for the explicit expression, see Ref.~\onlinecite{Smith1998}).
We found that the electrostatic energy from the cutoff procedure differs from the corresponding energy using the Ewald summation technique by only half a per mille.
This justifies the simpler cut-off procedure.

\subsection{\label{sec:Bulk crystal at room temperature}Bulk crystal at room temperature}

In the following, we investigate the equilibrium structure of the bulk crystal for the point quadrupole model  (see Sec.~\ref{sec:Point quadrupole approach}; quadrupole strengths: $Q^*=0, 0.5, 1, 1.5$) and the implicit electrostatics model (see Sec.~\ref{sec:Charged ring approach})
using trajectory data from the Molecular Dynamics simulations.
One important criteria to judge the performance of the different models are the unit cell parameters listed in Tab.~\ref{tab:unitcellparameters}.

\begin{widetext}

\begingroup
\squeezetable
\begin{table}[htb]
\begin{ruledtabular}
\caption{\label{tab:unitcellparameters}
Variation of unit cell parameters for the different pair potentials at room temperature.
The box lengths of the cell are denoted with $a$, $b$ and $c$ defined through the length of the unit cell vectors $\mathbf{a}$, $\mathbf{b}$ and $\mathbf{c}$. The angles between the unit cell vectors are denoted with $\alpha=\measuredangle(\mathbf{b},\mathbf{c})$, $\beta=\measuredangle(\mathbf{a},\mathbf{c})$ and $\gamma=\measuredangle(\mathbf{a},\mathbf{b})$.
The volume and the cohesion energy per particle are listed as well.
}
\begin{tabular}{ccccccccc}
Model&$a($Å$)$& $b($Å$)$&$c($Å$)$&$\alpha($°$)$ & $\beta($°$)$ & $\gamma($°$)$&$V($Å$^3)$&$E_{\text{coh}}($eV$)$\\
\hline
&&&&&&&\\
pq model $(Q^*\!=\!0)$ / vdW model&19.6444&3.6282&9.1430&90.02&117.74&89.98&576.78&1.5533\\
pq model $(Q^*\!=\!0.5)$&17.0045&4.2382&9.2411&90.00&113.66&90.00&610.04&1.7153\\
pq model $(Q^*\!=\!1)$&17.1237&4.0178&9.7228&90.00&116.52&90.00&598.53&2.9882\\
pq model $(Q^*\!=\!1.5)$&17.7104&3.7246&9.9783&89.62&122.77&86.83&552.08&6.4402\\
\hline
&&&&&&&\\
impl. electr. model&17.6142&4.5524&9.5722&90.00&112.91&90.01&707.03&0.9457\\
\hline
&&&&&&&\\
experiment~\cite{Echigo2007}&$16.094(9)$ & $4.690(3)$ & $10.049(8)$ & $90$ & $110.79(2)$ & $90$&$709.9(8)$ &/\\
experiment~\cite{Robertson1944,Robertson1945}&$16.119$ & $4.702$ & $10.102$ & $90$ & $110.9$ & $90$&$717.1$&$1.39-1.54$\\
database~\cite{Fawcett1966} ($E_{\text{coh}}$ from DFT-calculations~\cite{Sancho-Garcia2015})&$16.11$ & $4.70$ & $10.10$ & $90$ & $110.9$ & $90$&$714.43$ &1.378-1.783\\
\hline
\end{tabular}
\end{ruledtabular}
\end{table}
\endgroup

\end{widetext}

Inspecting the values for $a$, $b$, $c$, $\alpha$, $\beta$, $\gamma$ (for description see caption of Tab.~\ref{tab:unitcellparameters}), we find very good agreement of the implicit electrostatics model and the point quadrupole model for weak
strength ($Q^*=0.5$) with the corresponding experimental data.
The remaining models  ($Q^* \neq 0.5$) reveal slight deviations in $a$, $b$, $c$, and $\beta$.
Nonetheless, the crystal structures are in all cases monoclinic ($\alpha=\gamma=90\mathring{}$).
Concerning the cohesion energy predicted by the point quadrupole models, we note a significant increase with the quadrupole strength.
This stems from the overestimation of the binding energy for parallel displaced configurations (see Fig.~\ref{fig:differentq}).
The cohesion energy of the implicit electrostatics model is somewhat underestimated.

We further want to analyze the orientation of the corresponding molecules (named A and B) in the unit cell.
Therefor, we introduce the herringbone angle $\phi$, which marks the angle between both molecular orientation axes that point ``upwards'' along $\mathbf{b}$ (see Fig.~\ref{fig:unitcell}), and is defined as follows
 \begin{align}
 \phi=\arccos\left[\left<\sgn(\hat{\mathbf{u}}_{\rm A}\cdot \mathbf{b})\hat{\mathbf{u}}_{\rm A}\right> \cdot \left<\sgn(\hat{\mathbf{u}}_{\rm B}\cdot \mathbf{b}) \hat{\mathbf{u}}_{\rm B}\right>\right]\text{.}
\end{align}

In Table~\ref{tab:unitcellparameters melting} all molecular angles with respect to the body fixed unit cell are displayed.
Specifically, we present the enclosed angles of the molecule A's axis with each unit cell axis as performed in Ref.~\onlinecite{Robertson1945}.

\begingroup
\squeezetable
\begin{table}[htb]
\begin{ruledtabular}
\caption{\label{tab:unitcellparameters melting}
The angles of molecule A with the $\mathbf{a}$, $\mathbf{b}$ and its perpendicular axis $\mathbf{a}\times\mathbf{b}$ are denoted by $\chi_N^{\rm A}$, $\psi_N^{\rm A}$ and $\omega_N^{\rm A}$.
The notation is taken from Ref.~\onlinecite{Robertson1945}.
}

\begin{tabular}{lcccc}
Model&$\chi_N^{\rm A}($°$)$&$\psi_N^{\rm A}($°$)$&$\omega_N^{\rm A}($°$)$& $\phi($°$)$\\
\hline
&&&&\\
pq model $(Q^*\!=\!0)$ / vdW model&95.65&12.20&79.24&24.38\\
pq model $(Q^*\!=\!0.5)$&129.69&39.86&93.12&79.72\\
pq model $(Q^*\!=\!1)$&129.83&40.66&96.85&81.32\\
pq model $(Q^*\!=\!1.5)$&121.90&40.79&108.47&81.54\\
\hline
&&&&\\
impl. electr. model &130.23&40.31&92.30&80.63\\
\hline
&&&&\\
ideal crystal~\cite{Robertson1944}&133.7&43.7&89.6&87.35\\
\end{tabular}
\end{ruledtabular}
\end{table}

As expected, we observe a nematic phase for the van der Waals model (point quadrupole model with a zero quadrupole strength) reflected by a small value for the enclosed 
angle of the molecule A's axis with the $\mathbf{b}$-axis, called $\psi_N^{\rm A}$ and a small herringbone angle $\phi$.
This finding is in accord with our previous work.~\cite{firstpub2014}
The alignment of the molecules with respect to all unit cell axes ($\chi_N^{\rm A}$, $\psi_N^{\rm A}$, $\omega_N^{\rm A}$) is for both, the implicit electrostatics model and the point quadrupole model with weak strength ($Q^*=0.5$), in good agreement with the experimental values.
For all investigations (except for the nematic phase of the model without electrostatics), no significant change for the distance of neighboring columns was observed (corresponding distances are $8.7$\,Å and $9.8$\,Å).

\subsection{\label{sec:Melting of bulk crystal}Melting of bulk crystal}
We now turn to the investigation of structural changes of the bulk crystal [see Fig.~\ref{fig:unitcell}(b)] upon heating up the system.
Hereby, we expect the crystal to melt.
The temperature range considered covers $300\,{\rm K}\le T\le 1500\,{\rm K}$.
After every ${100\,{\rm K}}$ a different isothermal-isobaric simulation run is performed.

We further investigate $\phi$ as a function of temperature for the different models as presented in Fig.~\ref{fig:hbandvolume}(a).
The melting from crystalline to isotropic phases is also reflected in Fig.~\ref{fig:hbandvolume}(b) showing the third root of the volume as a function of the temperature.

\begin{figure}[htb]
     \includegraphics[width=8.5cm]{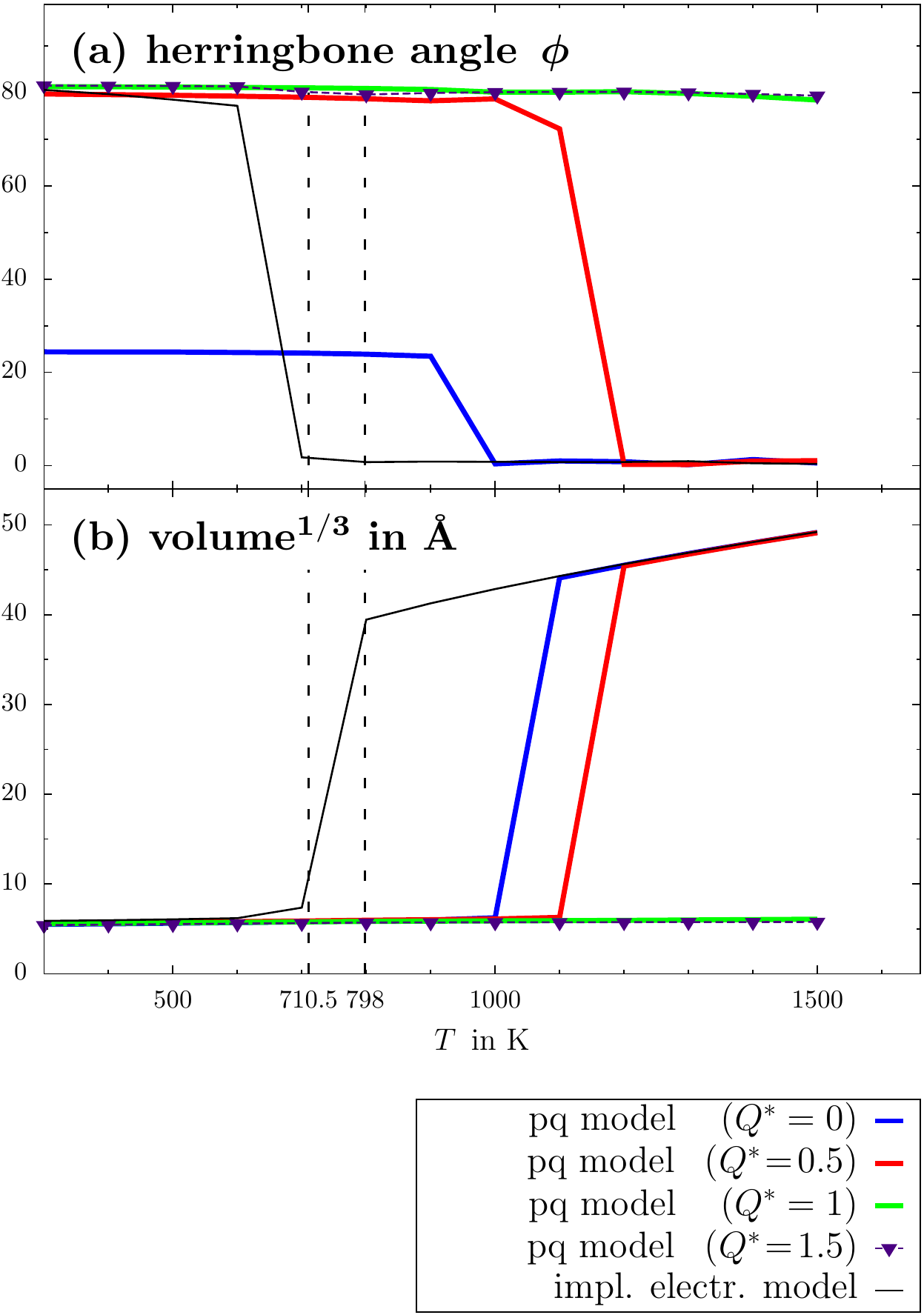}
  \caption{\label{fig:hbandvolume}(a) Herringbone angle between neighboring discs and (b) volume as function of the temperature. }
\end{figure}
When considering the reference temperatures of bulk coronene with respect to melting (${710.5 \,{\rm K}}$)~\cite{Schmidt2006} and boiling (${798 \,{\rm K}}$)~\cite{Schmidt2006}, we recognize
that coronene is liquid only in a narrow temperature band. In the implicit electrostatics model we observe a decay of the herringbone angle between $600$ and ${700 \,{\rm K}}$ towards zero indicating isotropic orientation.
At ${700 \,{\rm K}}$ the volume is significantly increased in comparison to ${600 \,{\rm K}}$, but highly raises with increasing temperature.
We take these as indications of a liquid phase appearing at around ${700 \,{\rm K}}$ and a gas phase at temperatures bigger than ${800 \,{\rm K}}$.
On the contrary to the implicit electrostatics model, the crystal structure for the point quadrupole model melts at significantly higher temperatures for all quadrupole strengths.
It seems obvious that the melting temperature increases with the cohesion energy, as discussed before (see Sec.~\ref{sec:Bulk crystal at room temperature}).

Finally, we want to analyze the crystalline order parallel to the plane spanned by $\mathbf{L}_1$ and $\mathbf{L}_3$ with the following correlation function~\cite{Andrienko2006}
\begin{align}
 g_{||}(h)&=\left<\frac{2}{V_{||}\,N\, \rho} \sum_{j}\sum_{k>j} f(h,\mathbf{\hat{n}}\cdot \mathbf{R}_{jk})\right>\text{,}\label{eqn:gpoa}
 \end{align}
where $\rho$ is the density and ${f(x,y)}$ equals unity for ${y \in \left( x -\frac{\Delta}{2}, x +\frac{\Delta}{2} \right]}$, otherwise ${f=0}$ (with $\Delta$ being the bin size).
The volume appearing in Eq.~\eqref{eqn:gpoa} is defined as ${V_{||}=\Delta\, \left| \mathbf{L}_1 \times \mathbf{L}_3 \right|}$.
For the implicit electrostatics model, we observe a crystalline order [see Fig.~\ref{fig:g_p}(a)], that continuously vanishes with rising temperature.
In contrast, the point quadrupole models do not exhibit this behavior [see Fig.~\ref{fig:g_p}(b)-(e)]. At a  quadrupole strength of $Q^*=1.5$ we do not see a positional order any more, while the orientational order still exists (see Fig.~\ref{fig:hbandvolume}).

 \begin{figure*}[htb]
  \centering \includegraphics*[width=13cm]{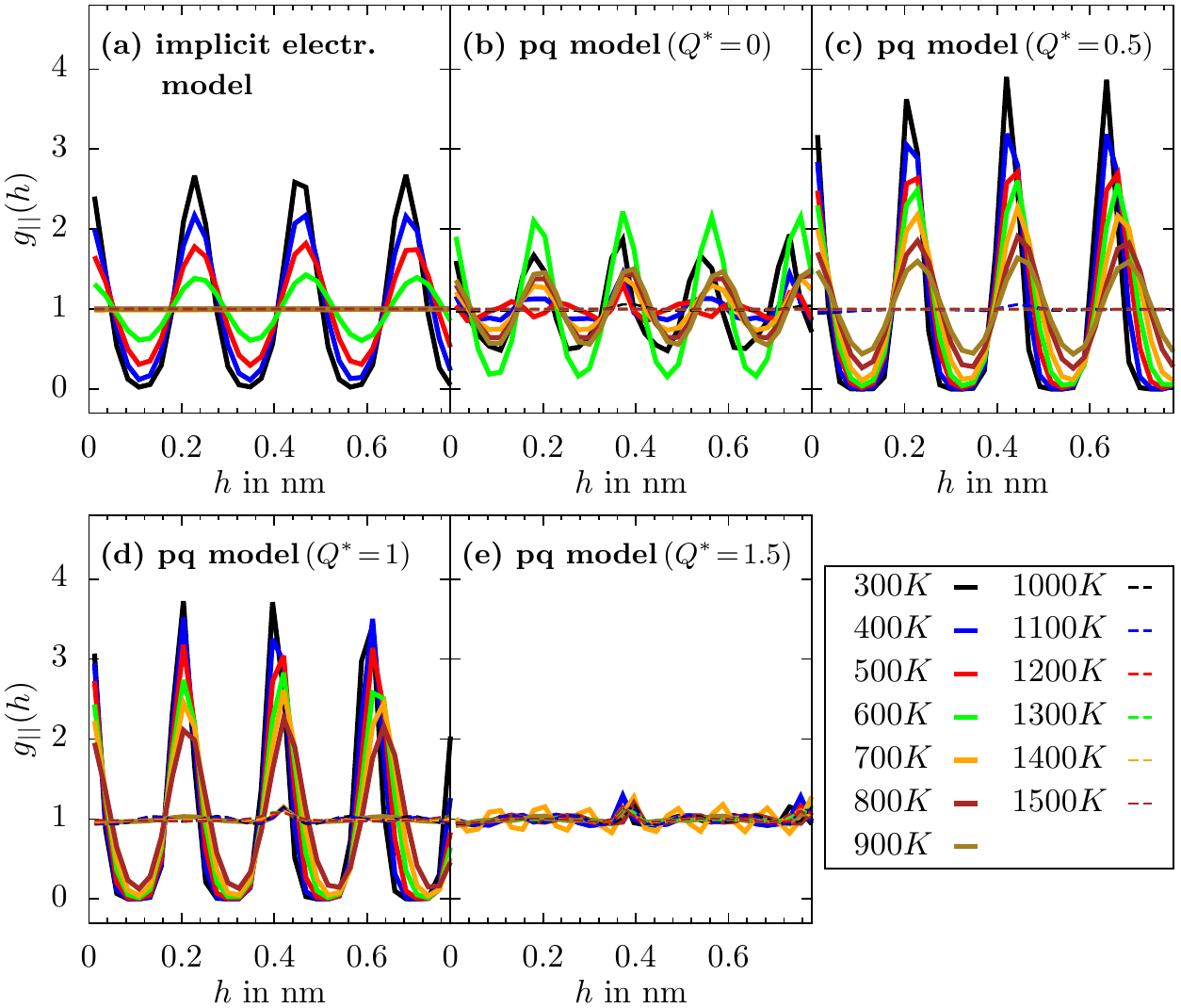}
  \caption{\label{fig:g_p}Temperature influence of pair correlation function along the $\mathbf{L}_1$-$\mathbf{L}_3$-plane for the implicit electrostatics model and the point quadrupole model for various strengths.}
\end{figure*}
To sum up, the transition temperature of the implicit electrostatics model can be suitably reproduced. Clearly it is not in full accordance with the experimental reference values.
Nonetheless, this approach marks a way to treat the relevant electrostatics in a rather simple model eligible for large-scale computer simulations.

\section{\label{sec:Conclusion}Conclusion}

In this study we presented different approaches to model the full pair interaction (with electrostatics) between coronene molecules, starting from a previously introduced van der Waals-like of model (see Sec.~\ref{sec:Treatment of the van der Waals interaction}) that neglected electrostatics.
Hereby, coronene serves as an exemplary discotic molecule with head-to-tail symmetry.
In the first approach (see Sec.~\ref{sec:Point quadrupole approach}), we aimed at extending this van der Waals model by a point quadrupole attached at the particles' center of mass along the particles' symmetry axis.
In the second approach (see Sec.~\ref{sec:Charged ring approach}) the van der Waals model was extended via electrostatic interactions stemming from charged rings.~\cite{Obolensky2007}
These ring interactions were used to reparametrize the van der Waals model into the so-called implicit electrostatics model.
In both electrostatic approaches, we treated the relevant electrostatics separate from the van der Waals interactions via a direct sum.

To assess the quality of modeling, unit cell parameters and structural quantities of the coronene bulk crystal were calculated with constant pressure and temperature Molecular Dynamics.
These results have been compared to their experimentally determined counterparts (see Sec.~\ref{sec:Many-particle simulations}).

Based on our data, we can conclude that the simple point quadrupole approach, although yielding the correct long-range decay gives unreliable results, even when the quadrupole strength is reduced to
a value related to a reasonable cohesion energy.

Moreover, for all considered quadrupole strengths, we observed melting temperatures far beyond the experimental values.
Nevertheless, a small point quadrupole leads to a stabilization of the herringbone structure due to an energetic preference.
In contrast, our second approach involving the implicit electrostatics model is able to stabilize the herringbone structure up to a melting point similar to experimental values.
With this second model, we also encountered a liquid phase as observed in the experiments.
As a conclusion, the implicit electrostatics model seems superior in reproducing structure and melting.
It is also more convenient from a computational perspective since it is just a reparametrized Gay-Berne-like model.
The crucial point for the success of the latter approach consists of choosing an extended charge distribution rather than a single point multipole.
Hereby the characteristics of the electrostatic potential in the near field are reproduced in a great extent.
Still, it would be interesting to investigate the influence of an atomic point charge distribution on the structural and melting properties of the coronene crystal.

Besides the temperature-independent electrostatics and the ring-charge assumption, a further underlying assumption in 
our work is the pairwise additivity of the many-molecule interactions.
The implications of this assumption were already investigated for crystalline benzene,~\cite{Evans1976} which is similar to coronene.
Further, it is worth mentioning that our temperature-dependent  approach does not take into account formation and breaking of covalent bonds, which may be an important processes for very high temperatures and pressures.~\cite{Jennings2010}

For future work, it would be rather interesting to use ab-initio simulations to calculate effective pair potentials for coronene and compare them with
our corresponding results for different temperatures.
Finally, we propose applicability of our presented method to a similar class of discotic molecules, e.g. benzene or hexabenzocoronene.

\begin{acknowledgments}
This work was supported by the Deutsche Forschungsgemeinschaft (DFG) within the framework of the CRC 951
(Project Nos. A7 and A1).
We thank Prof. M. Mazars for fruitful discussions concerning the Ewald summation technique.
\end{acknowledgments}

\appendix

\begin{widetext}
\section{\label{sec:Dimer configurations}Dimer configurations}
In Table~\ref{tab:dimerconfigs} we  present important examples of dimer configurations in terms of four reaction coordinates.
The reaction coordinates $R$, $a$,$b$ and $c$ are related to the molecular center of mass positions $\mathbf{R}_{\text{A}}$, $\mathbf{R}_{\text{B}}$ and the 
orientations $\mathbf{\hat{u}}_{\text{A}}$ and $\mathbf{\hat{u}}_{\text{B}}$ as follows
\begin{align}\label{eqn:reaction coordinates}
R&=\left\|\mathbf{R}_{\text{B}}-\mathbf{R}_{\text{A}}\right\| \text{,}\\
a&=\left\|\mathbf{\hat{u}}_{\text{A}}\cdot\mathbf{\hat{R}}\right\| \, \text{with} \, \mathbf{\hat{R}}= \frac{\mathbf{R}_{\text{B}}-\mathbf{R}_{\text{A}}}{\left\|\mathbf{R}_{\text{B}}-\mathbf{R}_{\text{A}}\right\|}\text{,}\\
b&=\left\|\mathbf{\hat{u}}_{\text{B}}\cdot\mathbf{\hat{R}}\right\| \text{,} \\
c&=\sgn(\mathbf{\hat{u}}_{\text{A}}\cdot\mathbf{\hat{R}})\,\sgn(\mathbf{\hat{u}}_{\text{B}}\cdot\mathbf{\hat{R}})\,\mathbf{\hat{u}}_{\text{A}}\cdot\mathbf{\hat{u}}_{\text{B}}\text{.}
\end{align}
These coordinates describe a molecular pair in the body-fixed frame, providing that the particles are uniaxial and have head-to-tail symmetry.
In addition $a$, $b$ and $c$ equally treat chiral dimer configurations.

\begingroup
\squeezetable
\begin{table}[htb]
\begin{ruledtabular}
\caption{\label{tab:dimerconfigs}
This table summarizes the interesting configurations with their corresponding values of the reaction coordinates
according to Ref.~\onlinecite{firstpub2014}.
}
\begin{tabular}{r|ccc|ccc|ccc|ccc|ccc|ccc|ccc|ccc}
&\multicolumn{3}{c|}{face-face}&\multicolumn{3}{c|}{parallel weakly}&\multicolumn{3}{c|}{parallel displaced}&\multicolumn{3}{c|}{T}&\multicolumn{3}{c|}{herringbone}
&\multicolumn{3}{c|}{V}&\multicolumn{3}{c|}{edge-edge}&\multicolumn{3}{c}{cross}\\
&\multicolumn{3}{c|}{\includegraphics[height=1.8em]{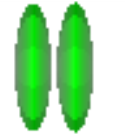}}&\multicolumn{3}{c|}{displaced \includegraphics[height=1.8em]{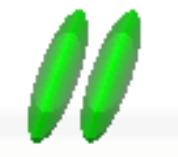}}&
\multicolumn{3}{c|}{\includegraphics[height=1.8em]{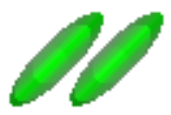}}&\multicolumn{3}{c|}{\includegraphics[height=1.8em]{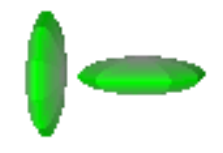}}&
\multicolumn{3}{c|}{\includegraphics[height=1.8em]{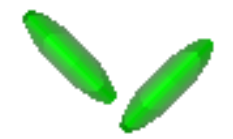}}&\multicolumn{3}{c|}{\includegraphics[height=1.8em]{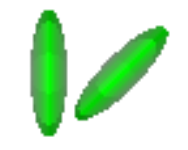}}&
\multicolumn{3}{c|}{\includegraphics[height=1.8em]{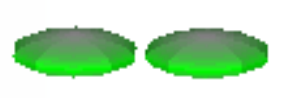}}&\multicolumn{3}{c}{\includegraphics[height=1.8em]{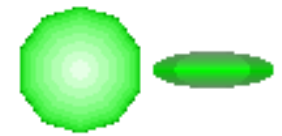}}\\
& a & b & c& a & b & c& a & b & c& a & b & c& a & b & c& a & b & c& a & b & c& a & b & c\\
\hline
&&&&&&&&&&&&&&&&&&&&&&&&\\
value & $1$ & $1$ & $1$ & $0.94$ & $0.94$ & $1$ & $1/\sqrt{2}$ & $1/\sqrt{2}$ & $1$ & $0$ & $1$ & $0$ & $0.8767$ & $0.481$& $0$ & $1$ & $1/\sqrt{2}$ & $1/\sqrt{2}$& $0$ & $0$ & $1$& $0$ & $0$ & $0$
\end{tabular}
\end{ruledtabular}
\end{table}
\endgroup

\FloatBarrier

\section{\label{sec:Ground state curves}Ground state curves}

In order to compare our results with ground state (GS) results in the literature, we consider in this Appendix a further model, which combines
the van der Waals potential pertaining to the atomistic level $U^{\rm GS}_{\rm vdW}$ (using generalized AMBER force field~\cite{Wang2004} without partial charges) at ${T=0\,{\rm K}}$ and
our ring-ring potential from Eq.~\eqref{eqn:ring-ring}.
Specifically,
\begin{align}
 U^{\rm GS}_{\rm direct}=U^{\rm GS}_{\rm vdW}+U_{\rm ring-ring}\text{.}
 \label{eqn:ground state direct model}
\end{align}
We term this potential ``ground state direct model''. In Fig.~\ref{fig:differentqmore} we present the ground state potential curves
$U^{\rm GS}_{\rm vdW}$ (without electrostatics) and $U^{\rm GS}_{\rm direct}$ [with ring-ring electrostatics from Eq.~\eqref{eqn:ring-ring}] using various configurations suitable for comparison with literature results.~\cite{Rapacioli2005,Zhao2008}
In addition, we add the potential curves showing the ground state analogon of the point quadrupole model, that is
\begin{align}
 U^{\rm GS}_{\rm pq}=U^{\rm GS}_{\rm vdW}(\mathbf{R},\mathbf{\hat{u}}_{\rm A},\mathbf{\hat{u}}_{\rm B})+\frac{1}{4\pi\varepsilon_0}\frac{3}{4} \frac{Q^2}{R^5}\cdot  K(\mathbf{\hat{R}},\mathbf{\hat{u}}_{\rm A},\mathbf{\hat{u}}_{\rm B})\text{.}
 \label{eqn:ground state qp model}
\end{align}
\begin{figure}[htb]
   \includegraphics*[width=14cm]{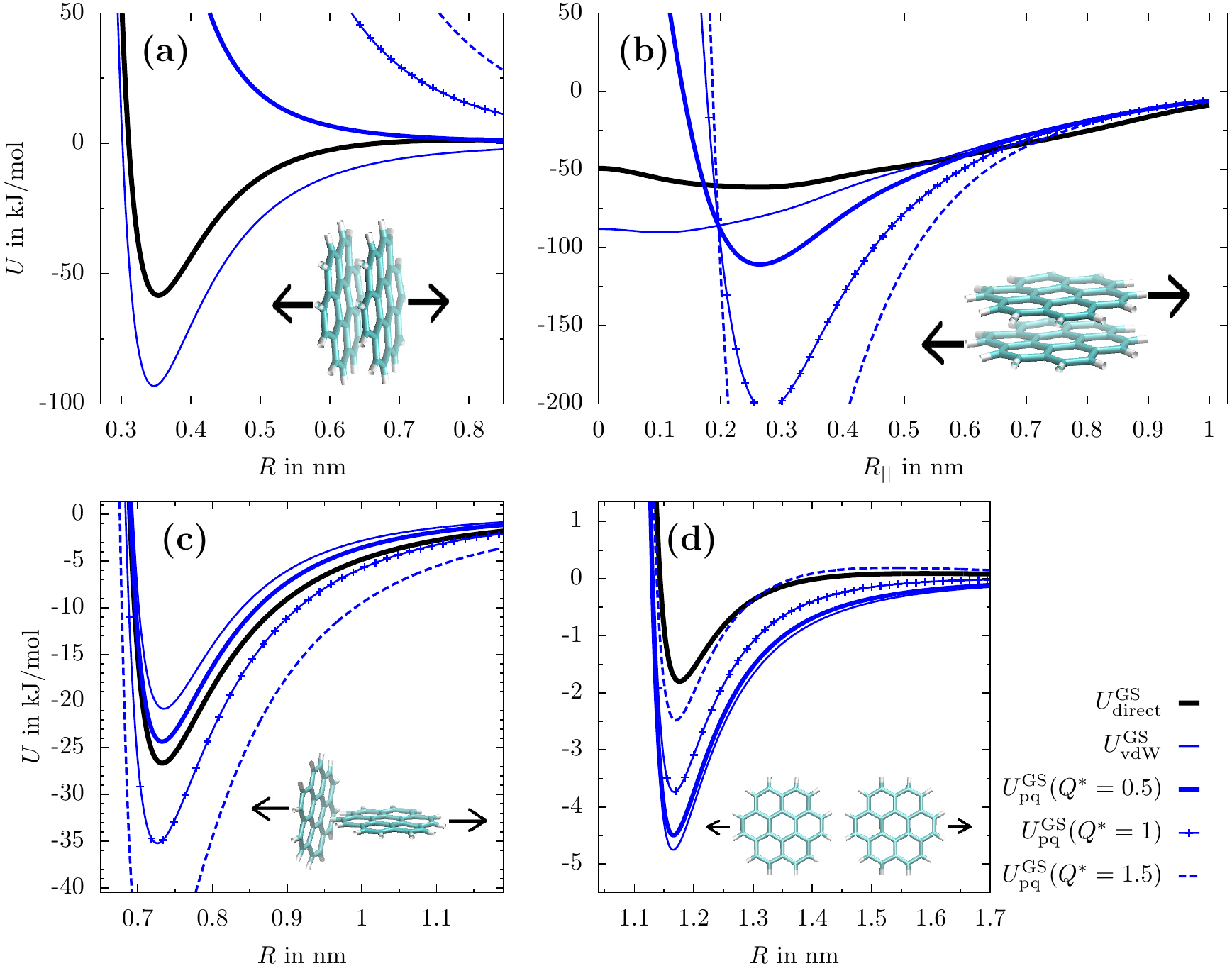}
    \caption{\label{fig:differentqmore} 
    Potential curves of the ground state models $U^{\rm GS}_{\rm direct}$ and $U^{\rm GS}_{\rm vdW}$, the direct model $U_{\rm direct}$, the implicit electrostatics model $U_{\rm implic.}$ and the point quadrupole model $U^{\rm GS}_{\rm qp}$ (reduced strengths: $0.5$, $1$, $1.5$) at ${T=300\,{\rm K}}$.
    (a) face-face configuration (with tooth-to-tooth setup); (b) parallel displaced configurations for a layer distance of $0.332$\,nm; (c) T configuration; (d) edge-edge configuration.}
 \end{figure}
\end{widetext}

We recognize that $U^{\rm GS}_{\rm direct}$ is characterized by similar values of the binding energy for the ff-configuration in Fig.~\ref{fig:differentqmore}(a) ($-58.31\,{\rm kJ}/{\rm mol}$ with tooth-to-tooth setup) and the pd-configuration in Fig.~\ref{fig:differentqmore}(b) ($-61.37\,{\rm kJ}/{\rm mol}$), consistent to what was observed in the work of Rapacioli et al~\cite{Rapacioli2005} (ff-configuration: $-94.9\,{\rm kJ}/{\rm mol}$, pd-configuration: $-93.66\,{\rm kJ}/{\rm mol}$).
However, the magnitudes of $U^{\rm GS}_{\rm direct}$ are only about two thirds of that obtained in Ref.~\onlinecite{Rapacioli2005}.
These differences stem from different charge distributions used by us and in Ref.~\onlinecite{Rapacioli2005}, and from the fact that our atomistic parameters are taken from the generalized AMBER force field~\cite{Wang2004} while in Ref.~\onlinecite{Rapacioli2005} parameters from van de Waal were used.
Still, we conclude that the curves of $U^{\rm GS}_{\rm direct}$ reveal far more lower binding energies than the ${300\,{\rm K}}$ curves of $U_{\rm direct}$ and are thus 
closer to the mentioned literature values.

For the sake of completeness, we also present curves for the t- and edge-edge configuration in Fig.~\ref{fig:differentqmore}(c) and ~\ref{fig:differentqmore}(d).
The general impact of the electrostatics is reflected by the difference between $U^{\rm GS}_{\rm direct}$ and $U^{\rm GS}_{\rm vdW}$ and is quite pronounced in all configurations.
As mentioned in Sec.~\ref{sec:Potential curves} the point quadrupole curves strongly affect the binding energies for all configurations.

\section{\label{sec:Temperature dependence}Temperature dependence}  

The temperature dependence of the face-face configuration is shown in Fig.~\ref{fig:ff temperature dependence} for the direct model and the ground state direct model [see Eq.~\eqref{eqn:ground state direct model}] for two different configurations (tooth-to-tooth and $30^{\circ}$-twisted).
\begin{figure}[htb]
   \includegraphics*[width=7.5cm]{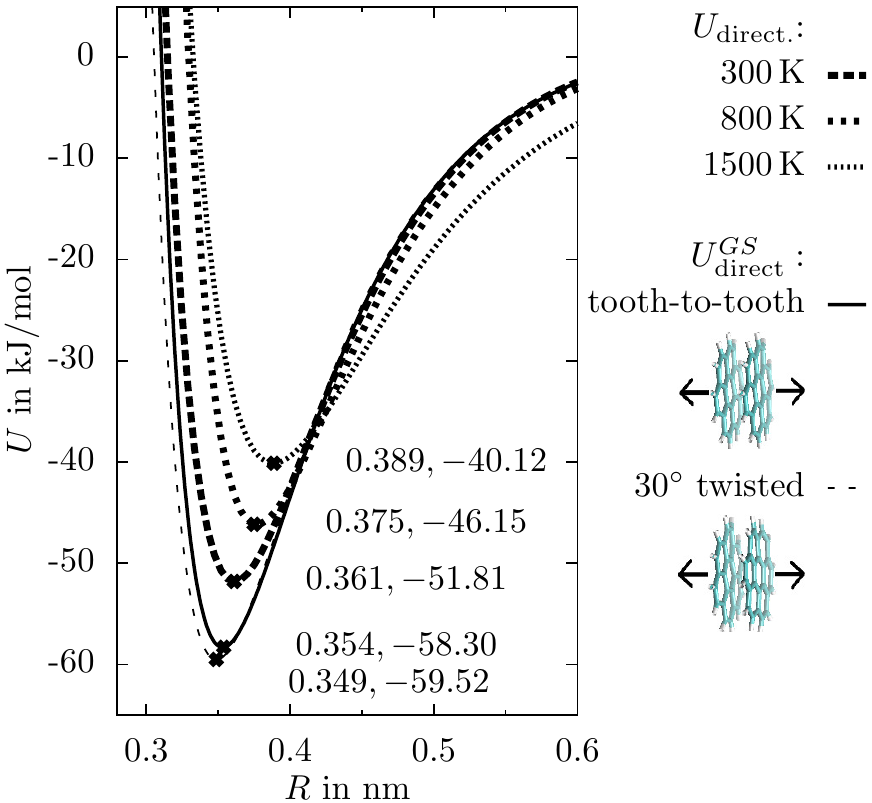}
    \caption{\label{fig:ff temperature dependence} Potential curves for the face-face configuration using the direct model at $300\,\mathrm{K}$, $800\,\mathrm{K}$ and $1500\,\mathrm{K}$. In addition two ground state configurations are added.}
 \end{figure}
  
 We can state that the  higher the temperature, the flatter the potential minimum and the larger the binding distance.
 The temperature dependence mainly stems from molecular bending modes as discussed in Ref.~\onlinecite{firstpub2014}.
 Its dependence due to axial averaging is rather small, reflected by the small difference of both ground state potentials.
 It is also worthwhile to mention that there is no influence of temperature at around ${R=0.42\,{\rm nm}}$, because
 at this point all potential curves intersect.
 
\FloatBarrier
\begin{widetext}
\section{Parametrizations}\label{sec:Parametrizations}
The following tables summarize the parameters used in Eq.~\eqref{eqn:UmGB} for the van der Waals model (Tab.~\ref{tab:Van der Waals model}) and the implicit electrostatics model (Tab.~\ref{tab:Implicit electrostatics model}) at different temperatures $T$.
\begingroup
\squeezetable
\begin{table}[htb]
\begin{ruledtabular}
\caption{\label{tab:Van der Waals model}Van der Waals parametrization}
\begin{tabular}{cccccccccccc}
$T(\mathrm{K})$&$\kappa$&$\chi'$&$\mu$&$\nu$&$\sigma_0(\mathrm{nm})$&$\epsilon_0(\mathrm{kJ}/\mathrm{mol})$&$d_{\rm w}$&$\gamma$&$\gamma'$&$\theta$&$\xi$\\
\hline
&&&&&&&&&&&\\
300&0.2885&-0.7895&1&1&1.0529&6.5481&0.3884&4&4&0.1592&-0.1967\\
400&0.2892&-0.7916&1&1&1.0603&6.3553&0.3869&4&4&0.2019&-0.1984\\
500&0.2899&-0.7939&1&1&1.0678&6.1614&0.3854&4&4&0.2478&-0.2002\\
600&0.2905&-0.7962&1&1&1.0752&5.9666&0.3839&4&4&0.2971&-0.2022\\
700&0.2912&-0.7986&1&1&1.0826&5.7707&0.3825&4&4&0.3503&-0.2043\\
800&0.2919&-0.8012&1&1&1.0900&5.5739&0.3811&4&4&0.4078&-0.2065\\
900&0.2916&-0.8024&1&1&1.0906&5.4425&0.3901&4&4&0.4426&-0.2063\\
1000&0.2914&-0.8037&1&1&1.0912&5.3113&0.3992&4&4&0.4790&-0.2060\\
1100&0.2911&-0.8051&1&1&1.0918&5.1802&0.4082&4&4&0.5171&-0.2057\\
1200&0.2909&-0.8064&1&1&1.0925&5.0494&0.4172&4&4&0.5571&-0.2055\\
1300&0.2907&-0.8079&1&1&1.0931&4.9187&0.4262&4&4&0.5991&-0.2052\\
1400&0.2904&-0.8094&1&1&1.0937&4.7882&0.4352&4&4&0.6434&-0.2049\\
1500&0.2902&-0.8109&1&1&1.0943&4.6579&0.4442&4&4&0.6900&-0.2046\\
\end{tabular}
\end{ruledtabular}
\end{table}
\squeezetable
\begin{table}[htb]
\begin{ruledtabular}
\caption{\label{tab:Implicit electrostatics model}Implicit electrostatics parametrization}
\begin{tabular}{cccccccccccc}
$T(\mathrm{K})$&$\kappa$&$\chi'$&$\mu$&$\nu$&$\sigma_0(\mathrm{nm})$&$\epsilon_0(\mathrm{kJ}/\mathrm{mol})$&$d_{\rm w}$&$\gamma$&$\gamma'$&$\theta$&$\xi$\\
\hline
&&&&&&&&&&&\\
300&0.2856&-0.7639&1&1&1.1026&5.3125&0.3440&4&4&1.4825&-0.3123\\
400&0.2867&-0.7635&1&1&1.1088&5.2244&0.3406&4&4&1.5003&0.3120\\
500&0.2877&-0.7631&1&1&1.1149&5.1354&0.3373&4&4&1.5196&-0.3117\\
600&0.2888&-0.7627&1&1&1.1211&5.0456&0.3340&4&4&1.5403&-0.3114\\
700&0.2898&-0.7623&1&1&1.1273&4.9548&0.3307&4&4&1.5626&-0.3111\\
800&0.2909&-0.7618&1&1&1.1335&4.8633&0.3275&4&4&1.5865&-0.3107\\
900&0.2905&-0.7622&1&1&1.1362&4.7541&0.3378&4&4&1.6406&-0.3111\\
1000&0.2902&-0.7626&1&1&1.1390&4.6452&0.3480&4&4&1.6969&-0.3115\\
1100&0.2899&-0.7631&1&1&1.1418&4.5365&0.3582&4&4&1.7558&-0.3119\\
1200&0.2896&-0.7635&1&1&1.1446&4.4280&0.3683&4&4&1.8173&-0.3123\\
1300&0.2892&-0.7639&1&1&1.1474&4.3197&0.3784&4&4&1.8816&-0.3127\\
1400&0.2889&-0.7644&1&1&1.1501&4.2117&0.3885&4&4&1.9491&-0.3132\\
1500&0.2886&-0.7649&1&1&1.1529&4.1038&0.3984&4&4&2.0198&-0.3137\\
\end{tabular}
\end{ruledtabular}
\end{table}
\endgroup
\end{widetext}

%

\end{document}